\documentclass[english,keywords,amsmath,amssymb,twocolumn]{revtex4}
\usepackage[T1]{fontenc}
\usepackage[latin1]{inputenc}
\usepackage{babel}
\usepackage{graphicx}
\usepackage{color}
\usepackage{bm}
\usepackage{longtable}
\usepackage{amsmath}
\usepackage{amsfonts}
\usepackage{amssymb}
\usepackage{multirow}

\begin{document}

\title{Swapping Intra-photon entanglement to Inter-photon entanglement using linear optical devices}
\author{Asmita Kumari}
\author{Abhishek Ghosh}
\author{Mohit Lal Bera}
\author{A. K. Pan \footnote{akp@nitp.ac.in}}
\affiliation{ National Institute Technology Patna, Ashok Rajhpath, Patna, Bihar 800005, India}
\begin{abstract}
We propose a curious protocol for swapping the \textit{intra-photon} entanglement between  path and polarization degrees of freedom of a single photon to \textit{inter-photon} entanglement between two distance photons which have never interacted. This is accomplished by using an experimental setup consisting of three suitable Mach-Zehnder interferometers along with number of beam splitters, polarization rotators and detectors. Using the same setup, we have also demonstrated an interesting quantum state transfer protocol, symmetric between Alice and Bob. Importantly, the Bell-basis discrimination is not required in both the swapping and state transfer protocols. Our proposal can be implemented using linear optical devices.
\end{abstract}

\maketitle
\section{introduction}
Quantum physics emerges as a surprising yet natural outgrowth of the revolutionary discoveries of physics during the first decade of twentieth century and has resulted in an extraordinary revision of our understanding of the microscopic world. Some quantum features can be exploited for information processing tasks. In recent decades a flurry of works have been performed, which includes storage and distribution of information in between non interacting system (for reviews, see \cite{rev}). Quantum entanglement is a fundamental resource for performing many information processing tasks including secret key distribution \cite{Bennet84} and dense coding \cite{Bennet92}. In 1993, Bennett and colleagues \cite{bennett93} put forwarded a path breaking protocol for transporting an unknown quantum state from one location to a spatially separated one - a protocol now widely known as quantum teleportation. A shared entangled states between the two parties and a classical communication channel are required to perform the quantum teleportation task. Right after this proposal, Bouwmeester \emph{et al}. \cite{Bouweester1997} and Boschi \emph{et al}. \cite{Boshi1998} experimentally implemented the teleportation protocol using photonic entangled state. Later, various other systems, such as atoms   \cite{Cirac1994,Riebe2007,Tang2010}, ions \cite{Olmschenk2009}, electrons \cite{De2006} and superconducting circuits \cite{Baur2012,Joo2016,Li2016} have been used for experimentally demonstrating teleportation and interesting extensions were subsequently proposed, specially those regarding the teleportation of more than one qubit \cite{Bouweester2000}.

By exploiting the notion of quantum teleportation a fascinating consequence emerges known as entanglement swapping \cite{Pan1998,zukowski93}. In a swapping protocol, the entanglement can be generated between two photons which have never interacted. If photon A entangled with photon B and C entangled with photon D, then the entanglement can be created between A and D, although they never interacted in the past. However, the photons B and C need to be interacted with each other. The swapping of entanglement has been extensively studied both theoretically \cite{zukowski93,Bose1998} and experimentally \cite{Pan1998,Jennewein2001,Ma2012,schmid}. It is worthwhile to mention here that both the teleportation and entanglement swapping protocols require the Bell basis discrimination which is practically a difficult task to achieve using linear optical instruments. A number of experiment have recently been conducted to perform the Bell basis analysis using linear optical devices \cite{calsamiglia,schuck,jennewein,grice,lee, zhou,wang}.
    
The primary aim of the present paper is to demonstrate an interesting entanglement swapping protocol so that the intra-photon entanglement between the two degrees of freedoms of single photon is swapped to the intra-photon entanglement between two spatially separated photons. Note that, the inter-photon entanglement is relatively more fragile than intra-photon one because the former is more prone to decoherence. In an interesting work \cite{adhikari2010}, the swapping of this kind was proposed. In this work, we use a different and elegant setup than that is used in \cite{adhikari2010} but similar to \cite{Lima16} to propose our entanglement swapping protocol. The same setup can be used to perform quantum state transfer which is technically different from the usual teleportation protocol. Both of our swapping and state transfer protocols do not require Bell-basis discrimination. Although our protocol is quite close in terms of the spirit of the original swapping protocol \cite{Pan1998,zukowski93}, but instead of using four photons, we use two photons and the inter-photon entanglement between path and polarization degrees of freedom of each of the photons. A suitable experimental setup involving three Mach-Zehnder interferometers (MZIs) and a few other linear optical devices are used to accomplish this task. Curiously, the photons have never interacted with each other during the whole process of swapping and state transfer. 

The paper is organized as follows. In Section II, we propose an experimental setup of the entanglement swapping protocol by using simple linear optical devices which allows to swap a path-polarization intra-photon entanglement of single photon onto the polarization-polarization or path-path intra-photon entanglement between two spatially separated photons. Using the same setup, we demonstrate a curious quantum state transfer protocol in Section III. We provide a brief summary of our results in Section IV. 
\section{Entanglement swapping protocol}
Our experimental setup consists of three suitable $MZIs$ where  $MZ_{1}$  and $MZ_{3}$ belong to Alice and Bob respectively, and the third interferometer $MZ_2$ is shared by both as shown in the Figure 1. The entire setup consists of five $50:50$ beam splitters, five polarizing beam splitters, five polarization rotators, eight detectors and two mirrors are denoted by $BS_{i}(i=1,2...5)$, $PBS_{j}(j=1,2...5)$, $PR_{k} (k=1,2,3,4,5)$, $D_{l} (l=1,2..8)$ and $M_{m} (m=1,2)$ respectively.

This arrangement can be considered as a chained Hardy setup \cite{hardy}. The well-known Hardy setup was originally proposed for demonstrating the non-locality without inequalities. It uses two MZIs, one with electron and other with positron, coupled through a common beam splitter. The positron and electron annihilate if they simultaneously pass through that common beam splitter. This is crucial to produce the non-maximally entangled state required for demonstrating Hardy non-locality. Our setup (Figure 1) is a chained Hardy setups in the sense that $MZ_1$ and $MZ_2$ share the $BS_1$, and   $MZ_2$ and $MZ_3$ share the $BS_2$. If electrons pass through the $MZ_1$ and $MZ_3$ and positrons pass through $MZ_2$, then electrons and positrons annihilate at $BS_1$ and $BS_2$. 

In our setup, we use three indistinguishable photons for the implementation of our protocol in which an effect similar to annihilation at $BS_1$ and $BS_2$ is necessary for producing a suitable entangled state required for our purpose. For the case of photons, similar effect of annihilation of positron and electron can be achieved through the bunching of indistinguishable photons at $BS_1$ and $BS_2$. This effect has been extensively discussed in the literature (see, for example,\cite{Hong1987,Pan2012,irvine2005}). In particular, in \cite{irvine2005}, the Hardy paradox is experimentally tested by using the two indistinguishable photons and their bunching effect. Here in Figure 1 we assume that three photons which are indistinguishable. One may also consider that they come from same source and incident on $PBS_1$, $PBS_2$ and $PBS_3$. 

We note here that experiments \cite{yurk46,yurk68} have also been performed for testing Bell's theorem and EPR paradox by using independent particles. In \cite{yurk46}, the violation of local realism is demonstrated using the particles from independent sources and knowledge of  their generation can be ignored if they are indistinguishable. The indistinguishability of particles was then used as a resource for demonstrating the violation of local realism. In contrast, in our scheme two indistinguishable photons incident on $MZ_1$ and $MZ_2$ bunch at $BS_1$ which in turn  produces non-maximally entanglement between  the degrees of freedom of photons incidenting on  $MZ_1$ and $MZ_2$. Similarly arguments can be made for photons on $MZ_2$ and $MZ_3$.  In our protocol, the intra-photon entanglements of photon in $MZ_1$ and photon in $MZ_3$  is swapped to the inter-photon entanglement between polarization degrees of freedom of the photons incident on $MZ_1$ and $MZ_3$. Importantly, the photons in $MZ_1$ and $MZ_3$ have never interacted (in contrast to  \cite{yurk46,yurk68} ) during the whole process of entanglement swapping and state transfer process. The indistinguishablity of photons plays an important role but the swapping of entanglement protocol demonstrated here cannot be described as the use of indistinguishablity of photons as resource.  
 
The task of our protocol is to generate a polarization-polarization entangled state between the photons entering $MZ_1$ and $MZ_3$ respectively while ensuring that they never interact. Further, our goal is to transfer the polarization state of Alice to Bob or  polarization state of Bob to Alice. Let $a $, $b $ and $c$ are input modes of photons incidenting on  $MZ_1$, $MZ_2$ and $MZ_3$ respectively. Then the initial state of three photons are given by $|\psi_1\rangle = \hat{a}^{\dagger}_{H}|0 \rangle$, $|\psi_2 \rangle =\hat{b}^{\dagger}_{H} |0 \rangle$ and $|\psi_3 \rangle = \hat{c}^{\dagger}_{H}|0 \rangle$ respectively, where  $|0 \rangle$ is the vacuum state. For our purpose, we rotate the polarization by using two polarization rotators ($PR_1$ and $PR_2$) along the modes $a$ and $c$ before $MZ_1$ and $MZ_3$ respectively. The action of $PR_1$ transform $\hat{a}^{\dagger}_{H}|0 \rangle$ to $(\alpha \hat{a}^{\dagger}_{H}+ \beta \hat{a}^{\dagger}_{V})|0 \rangle $. Similarly the action of  $PR_2$ transform $ \hat{c}^{\dagger}_{H}|0 \rangle$ to $(\gamma \hat{c}^{\dagger}_{ H} +  \delta \hat{c}^{\dagger}_{V}) |0 \rangle$. Here $H$ and $V$ represents the horizontal and vertical polarization mode and $\alpha, \beta, \gamma$ and $\delta$ are the constants satisfying the normalization condition $|\alpha|^2 + |\beta|^2=1$ and $|\gamma|^2 + |\delta|^2=1$. 
  
Now, if  $a$ and $p$ are the output modes of the beam splitters $PBS_1$, the state of photon in $MZ_1$ after passing through $PBS_1$ transformed to an intra-photon entanglement between spatial mode and polarization can be written as $(\alpha \hat{a}^{\dagger}_{H} + i \beta \hat{p}^{\dagger}_{V})|0 \rangle $. Similarly the state of the photon in $MZ_2$ transformed to $ \frac{1 }{\sqrt{2}} (\hat{b}^{\dagger}_{ H} + i  \hat{q}^{\dagger}_{V})|0 \rangle $ after passing through $PBS_2$ having output mode $b$ and $q$. The state of the photon in $MZ_3$ after passing through $PBS_3$ transformed to an intra-photon entanglement $(\gamma \hat{c}^{\dagger}_{ H} + i \delta \hat{r}^{\dagger}_{V})|0 \rangle $, where $c$ and $r$ are the output modes of $PBS_3$. 

 Hence, the total state of the three photons emerging from $PBS_1$, $PBS_2$ and $PBS_3$ is given by
\begin{eqnarray}
\label{com}
\nonumber
|\Psi\rangle&=&\frac{1 }{\sqrt{2}} \bigg[(\alpha \hat{a}^{\dagger}_{ H} + i \beta \hat{p}^{\dagger}_{V})(\hat{b}^{\dagger}_{ H} + i  \hat{q}^{\dagger}_{V}) \\ && (\gamma \hat{c}^{\dagger}_{ H} + i \delta \hat{r}^{\dagger}_{V})|0 \rangle \bigg] 
\end{eqnarray}

Next, for understanding the operations of $M_1$, $BS_1$, $BS_2$ and $M_2$ on photons let us rearrange Eq.($\ref{com}$) in the following way
\begin{subequations}
	\begin{eqnarray}
		\label{rcom1}
	| \Psi\rangle = \frac{1}{\sqrt{2}}\big[ 
	&-& \beta \hat{p}^{\dagger}_{V}  \hat{q}^{\dagger}_{V}  \big (\gamma \hat{c}^{\dagger}_{ H} + i \delta \hat{r}^{\dagger}_{V} \big)\\
	\label{rcom2}
	&+& \gamma \big(\alpha \hat{a}^{\dagger}_{ H} + i \beta \hat{p}^{\dagger}_{V}\big)  
	\hat{b}^{\dagger}_{ H} \hat{c}^{\dagger}_{ H}   
	\\ 
	\label{rcom3}
	&+& i \delta \big(\alpha \hat{a}^{\dagger}_{ H} + i \beta \hat{p}^{\dagger}_{V} \big) \hat{b}^{\dagger}_{ H}  \hat{r}^{\dagger}_{V}
	\\
	\label{rcom4} 
	&+& i \alpha \hat{a}^{\dagger}_{ H} \hat{q}^{\dagger}_{V}  \big(\gamma \hat{c}^{\dagger}_{ H} + i \delta \hat{r}^{\dagger}_{V}) \big]|0 \rangle 
	\end{eqnarray}
\end{subequations}
\begin{widetext}
\begin{center}
\begin{figure}[ht]
  \includegraphics[width=8 cm,height=11 cm]{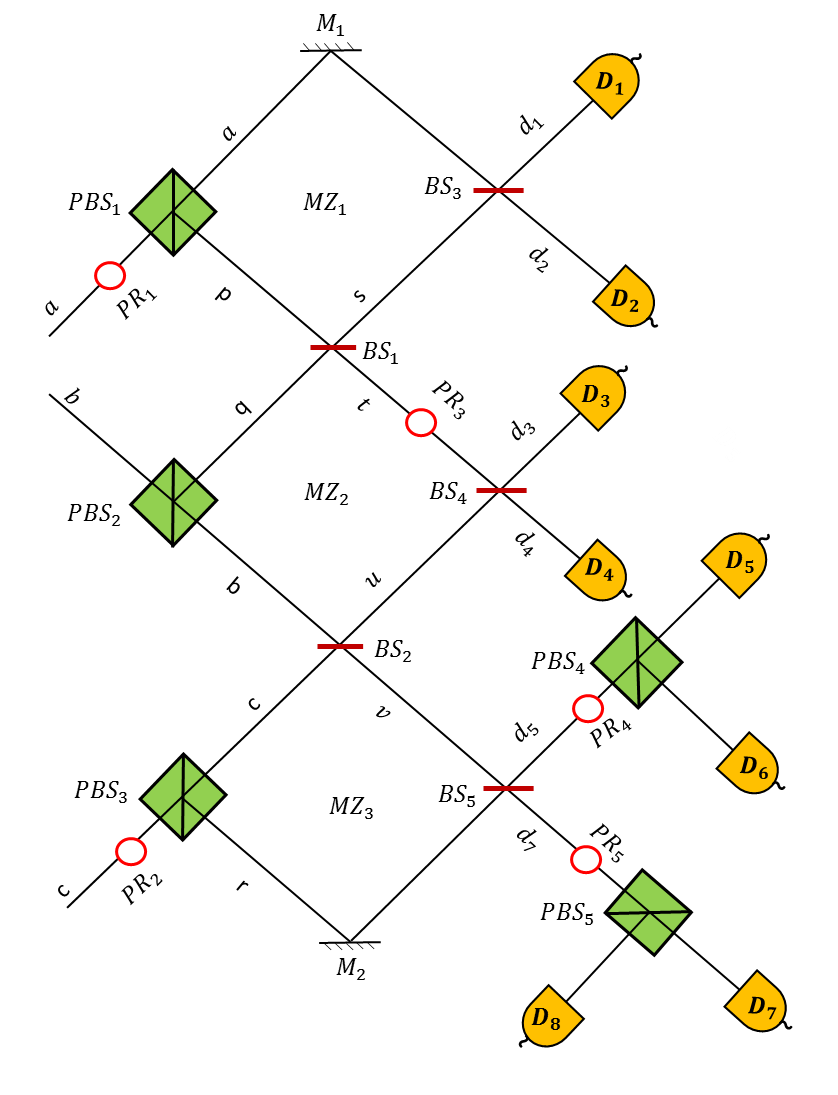}
 \centering
  \caption{(color online) The setup for implementing the swapping of intra-photon path-polarization entanglement of each of the photons in  $MZ_1$ and  $MZ_3$ to inter-photon  polarization -polarization  entanglement between the photons in  $MZ_1$ and  $MZ_3$  and for transferring polarization state of photon in  $MZ_1$ to photon in  $MZ_3$ (Details are given in the text).}
  \label{fig:1}
\end{figure}
\end{center}
\end{widetext}
In Eq.($\ref{rcom1}$), the special modes $p$ and $q$ coming from $PBS_1$ and $PBS_2$ respectively serve as input modes of  $BS_1$ (central beam splitter of $MZ_1$ and $MZ_2$) with same polarization $V$, results in bunching effect at $BS_1$ (similar to the annihilation in the case of electron and positron in Hardy's original paper). If $s$ and $t$ are the output modes of $BS_1$, then we have 
\begin{eqnarray}
\nonumber
   \hat{p}^{\dagger}_{V}  \hat{q}^{\dagger}_{V} |0 \rangle  & \rightarrow &\frac{1}{2}\left( \hat{s}^{\dagger}_{ V} + i  \hat{t}^{\dagger}_{V}\right)\left( i \hat{s}^{\dagger}_{V} +  \hat{t}^{\dagger}_{V}\right)|0 \rangle 
\end{eqnarray}
 Since photons are bosons and bosonic  creation operators commute, i.e., $\hat{s}^{\dagger}_{ V} \hat{t}^{\dagger}_{V} = \hat{t}^{\dagger}_{V} \hat{s}^{\dagger}_{ V} $, then the bunching effect at  $BS_1$  gives
\begin{eqnarray}
    \hat{p}^{\dagger}_{V}  \hat{q}^{\dagger}_{V} |0 \rangle  &\rightarrow& \frac{i}{\sqrt{2}} \left( \hat{s}^{\dagger 2}_{ V} + \hat{t}^{\dagger 2}_{V}\right)|0 \rangle 
\end{eqnarray}
 Similarly in Eq.($\ref{rcom2}$)  the special modes $b$ and $c$ coming from $PBS_2$ and $PBS_3$ respectively serve as input modes of  $BS_2$ with same polarization mode $H$  results in bunching effect at $BS_2$.  If $u$ and $v$ are the output modes of $BS_2$, then, the bunching effect provides
\begin{eqnarray}
\nonumber
 \hat{b}^{\dagger}_{ H} \hat{c}^{\dagger}_{ H} & \rightarrow& \frac{1}{2}\left( \hat{u}^{\dagger}_{ H} + i  \hat{v}^{\dagger}_{H}\right)\left( i \hat{u}^{\dagger}_{H} +  \hat{v}^{\dagger}_{H}\right)|0 \rangle \\
  &\rightarrow& \frac{i}{\sqrt{2}} \left( \hat{u}^{\dagger 2}_{ H}  +  \hat{v}^{\dagger 2}_{H}\right)|0 \rangle
\end{eqnarray}

Hence, the bunching effect excludes the possibility of detecting photons leaving each interferometer simultaneously and consequently coincidence clicks for the terms in Eqs.(\ref{rcom1}-\ref{rcom2}) are absent.

Next, the term $ \hat{a}^{\dagger}_{ H} \hat{b}^{\dagger}_{ H}  \hat{r}^{\dagger}_{V}|0\rangle$ in Eq.($\ref{rcom3}$) got phase shift of $-i$ due to three reflections at $M_1$, $BS_2$ and $M_2$ respectively. Due to the transmission of photon in spatial mode $b$ at $BS_2$ the amplitude gets reduced by the factor of $1/\sqrt{2}$. On the other hand the term $\hat{p}^{\dagger}_{V} \hat{b}^{\dagger}_{ H}  \hat{r}^{\dagger}_{V}|0 \rangle$  in Eq.($\ref{rcom3}$) got phase shift of $-i$ due to three reflections at $BS_1$, $BS_2$ and $M_2$ and, the amplitude gets reduced by the factor of 1/2 due to transmissions of  photons in spatial modes $ p $ and $b$ at $BS_1$ and $BS_2$ respectively. Hence the terms in Eq.($\ref{rcom3}$) evolves to
\begin{eqnarray}
  \label{t3}
  & i \delta \big(\alpha \hat{a}^{\dagger}_{ H} + i \beta \hat{p}^{\dagger}_{V} \big) \hat{b}^{\dagger}_{ H}  \hat{r}^{\dagger}_{V}|0 \rangle \\  \nonumber
  & \rightarrow \frac{\delta}{2}\big[ \sqrt{2} \alpha   \hat{a}^{\dagger}_{ H} \hat{u}^{\dagger}_{ H}  \hat{r}^{\dagger}_{V}   +i \beta  \hat{s}^{\dagger}_{V} \hat{u}^{\dagger}_{ H}  \hat{r}^{\dagger}_{V} \big]|0 \rangle& 
\end{eqnarray}
Similarly the term $\hat{a}^{\dagger}_{ H} \hat{q}^{\dagger}_{V} \hat{c}^{\dagger}_{ H}|0 \rangle $ in Eq.(\ref{rcom4}) got phase shift of $-i$ after three reflections at $M_1$, $BS_1$ and $BS_2$. Due to transmissions of photons in spatial mode $ q$ and $c$ at $BS_1$ and $BS_2$ respectively, the overall amplitude gets reduced by the factor of $1/2$. On the other hand the term $ \hat{a}^{\dagger}_{ H} \hat{q}^{\dagger}_{V} \hat{r}^{\dagger}_{V}) |0 \rangle $ in Eq.(\ref{rcom4}) shifted by the phase of $-i$ due to three reflections at $M_1$, $BS_1$ and $M_2$, and the amplitude of this term is reduced by $1/\sqrt{2}$ due to transmission of photon with the spatial mode $q$ at $BS_1$. The terms of Eq.(\ref{rcom4}) after passing through $M_1$, $BS_1$, $BS_2$ and $M_2$ evolves to 
\begin{eqnarray}
\label{t4}
& i \alpha \hat{a}^{\dagger}_{ H} \hat{q}^{\dagger}_{V}  \big(\gamma \hat{c}^{\dagger}_{ H} + i \delta \hat{r}^{\dagger}_{V}) |0 \rangle   \\  \nonumber
& \rightarrow \frac{\alpha}{2}\big[  \gamma \hat{a}^{\dagger}_{ H} \hat{t}^{\dagger}_{V} \hat{v}^{\dagger}_{ H}  +i \sqrt{2}  \delta  \hat{a}^{\dagger}_{ H} \hat{t}^{\dagger}_{V} \hat{r}^{\dagger}_{V}  \big]|0 \rangle&
\end{eqnarray}
Finally, the state of the three photons after the operation at $M_1$, $BS_1$, $BS_2$ and $M_2$ is given by
\begin{eqnarray}
\nonumber
| \Psi\rangle &=& N_1 \big[ \sqrt{2} \alpha \delta  \hat{a}^{\dagger}_{ H} \hat{u}^{\dagger}_{ H}  \hat{r}^{\dagger}_{V}   +i \beta \delta \hat{s}^{\dagger}_{V} \hat{u}^{\dagger}_{ H}  \hat{r}^{\dagger}_{V} \\  \nonumber 
& + & \alpha \gamma \hat{a}^{\dagger}_{ H} \hat{t}^{\dagger}_{V} \hat{v}^{\dagger}_{ H}  +i \sqrt{2} \alpha \delta  \hat{a}^{\dagger}_{ H} \hat{t}^{\dagger}_{V} \hat{r}^{\dagger}_{V}  \big]|0 \rangle
\end{eqnarray}
where $ N_{1} =(\alpha^{2} \gamma^{2} + 4\alpha^{2} \delta^{2} + \beta^{2}\delta^{2})^{-1/2 }$ is normalization constant. Using the polarization rotator $PR_3$ before $BS_4$ we flip polarization of  photon as shown in Figure 1. The final state is given by
\begin{eqnarray}
\nonumber
| \Psi_1\rangle &=&  N_1 \big[ \sqrt{2} \alpha \delta  \hat{a}^{\dagger}_{ H} \hat{u}^{\dagger}_{ H}  \hat{r}^{\dagger}_{V}   +i \beta \delta \hat{s}^{\dagger}_{V} \hat{u}^{\dagger}_{ H}  \hat{r}^{\dagger}_{V} \\   
& + & \alpha \gamma \hat{a}^{\dagger}_{ H} \hat{t}^{\dagger}_{H} \hat{v}^{\dagger}_{ H}  +i \sqrt{2} \alpha \delta  \hat{a}^{\dagger}_{ H} \hat{t}^{\dagger}_{H} \hat{r}^{\dagger}_{V}  \big]|0 \rangle
\end{eqnarray}

  Let us now consider two cases:
  
   (i) If $d_3$ and $d_4$ are the output modes of  $BS_{4}$, the state of the  photon after $BS_{4}$ $ d^{\dagger}_{4 H}|0 \rangle= \left(-i \hat{u}^{\dagger}_{ H} + \hat{t}^{\dagger}_{H}\right)|0 \rangle / \sqrt{2}$  results in a detection in $D_4 $.
   
   (ii) The state of the photon after $BS_{4}$  $d^{\dagger}_{3 H}|0 \rangle= \left(i \hat{u}^{\dagger}_{ H} + \hat{t}^{\dagger}_{H}\right)|0 \rangle / \sqrt{2}$ results in a different detector at $D_3 $. 
   
    In case (i), we end up with a four-qubit GHZ type entangled state of  spatial mode and polarization of the photons involving $MZ_1$ and $MZ_3$ can then be written as,
 
 \begin{eqnarray}
  \label{ent1}
 |\Psi_{2} \rangle &=&N_2( \beta \delta \hat{s}^{\dagger}_{V}   \hat{r}^{\dagger}_{V} + \alpha \gamma \hat{a}^{\dagger}_{ H} \hat{v}^{\dagger}_{ H})]|0 \rangle
 \end{eqnarray}
 where, $ N_{2} = (\alpha^{2}\gamma^{2} + \beta^2 \delta^2)^{-1/2}$. 
 
 We thus prepared an entangled state between the four degrees of freedoms between spatial modes and polariozations corresponding to the photons in $MZ_1$ and $MZ_3$ by introducing constrains in spatial modes and using a suitable projective measurement on the final output modes $d_3$ and $d_4$ of $MZ_2$. It is to be noted that during the whole process, the photons in $MZ_1$ and $MZ_3$ have never interacted with each other.

Similarly, for the case (ii), the resulting reduced state of the photons in $MZ_1$ and $MZ_3$ can be written as

\begin{eqnarray}
\label{ent}
|\Psi_2'\rangle &=& N'_{2} \big[ \alpha \gamma \hat{a}^{\dagger}_{ H} \hat{v}^{\dagger}_{ H} -  \beta \delta \hat{s}^{\dagger}_{V}   \hat{r}^{\dagger}_{V} \\  \nonumber
&& +  2 \sqrt{2} i \alpha \delta  \hat{a}^{\dagger}_{ H}  \hat{r}^{\dagger}_{V} \big]|0 \rangle
 \end{eqnarray}
where $N'_{2}=(\alpha^{2}\gamma^{2} + \beta^2 \delta^2 +8\alpha^2 \delta^2)^{-1/2}$. We do not further use the state in Eq.(\ref{ent}) in this paper.\\

In order to achieve the polarization-polarization entanglement between the photons in $MZ_1$ and $MZ_3$, we need to invoke a suitable disentangling process which again requires no direct interaction between the photons in $MZ_{1}$ and $MZ_{3}$. For this, we consider relations of input-output creation operator at the beam splitter $ BS_{3}$ as, $\hat{a}^{\dagger}_{H} = (\hat{d}^{\dagger}_{1H}+ i \hat{d}^{\dagger}_{2H}) / \sqrt{2} $  and $\hat{s}^{\dagger}_{V} = (i \hat{d}^{\dagger}_{1V}+  \hat{d}^{\dagger}_{2V}) / \sqrt{2} $. The state (Eq.(\ref{ent1})) after $BS_{3}$ can then be written as,
\begin{eqnarray}	
	|\Psi_{3} \rangle &=&\frac{N_2}{\sqrt{2}}(\beta \delta(i \hat{d}^{\dagger}_{1V}+  \hat{d}^{\dagger}_{2V})\hat{r}^{\dagger}_{V} \\  \nonumber
&+& \alpha \gamma(\hat{d}^{\dagger}_{1H}+ i \hat{d}^{\dagger}_{2H})\hat{v}^{\dagger}_{H})]|0 \rangle
\end{eqnarray}

Similarly, if at the beam splitter $BS_{5} $ relation between input-output creation-operator is $ \hat{v}^{\dagger}_{H}= (\hat{d}^{\dagger}_{5H}+ i \hat{d}^{\dagger}_{7H})/ \sqrt{2} $ and $\hat{r}^{\dagger}_{V} = (i \hat{d}^{\dagger}_{5V}+  \hat{d}^{\dagger}_{7V})/ \sqrt{2}$, then the joint state of the photons  in $MZ_1$ and $MZ_3$ after $BS_{3}$ and $BS_{5}$ becomes, 
\begin{eqnarray}
\label{joint}
\nonumber
	|\Psi_{4} \rangle &=&\frac{N_2}{2}[(\alpha \gamma \hat{d}^{\dagger}_{1H}\hat{d}^{\dagger}_{5H}-\beta \delta \hat{d}^{\dagger}_{1V}\hat{d}^{\dagger}_{5V} )\\ 
	\nonumber
	&&+i( \alpha \gamma \hat{d}^{\dagger}_{2H}\hat{d}^{\dagger}_{5H}+\beta \delta \hat{d}^{\dagger}_{2V}\hat{d}^{\dagger}_{5V})\\ 
	\nonumber
	&&+i (\alpha \gamma \hat{d}^{\dagger}_{1H}\hat{d}^{\dagger}_{7H}+\beta \delta \hat{d}^{\dagger}_{1V}\hat{d}^{\dagger}_{7V})\\ 
	&&-( \alpha \gamma \hat{d}^{\dagger}_{2H}\hat{d}^{\dagger}_{7H}-\beta \delta \hat{d}^{\dagger}_{2V}\hat{d}^{\dagger}_{7V})]|0 \rangle
\end{eqnarray}	
Depending on a suitable joint path measurement chosen by Alice and Bob the following  polarization-polarization inter-photon entangled state
\begin{eqnarray}
 |\Psi_{AB}\rangle = N_3 (\alpha \gamma \hat{d}^{\dagger}_{1H}\hat{d}^{\dagger}_{5H}-\beta \delta \hat{d}^{\dagger}_{1V}\hat{d}^{\dagger}_{5V})|0 \rangle
 \end{eqnarray}
  can be generated, where $N_3= (\alpha^2 \gamma^2 +\beta^2 \delta^2)^{-1/2}$. When Alice and Bob choose $d_{2}$ and $d_{5}$ respectively, an additional gate operation  $\hat{\sigma_{z}}$ is required for obtaining the entangled state $|\Psi_{AB}\rangle$. Similarly for the case when Alice and Bob choose  $d_{1}$ and $d_{7}$ respectively. If we take $\alpha=\beta=\gamma=\delta=1/\sqrt{2}$, the state $|\Psi_{AB}\rangle$ becomes maximally entangled.

Hence, using our setup we have generated a polarization-polarization entanglement between the photons in $MZ_1$ and $MZ_3$ even when they have never interacted with each other. It is important to note that, both the photons contain an intra-photon path-polarization entanglement that is swapped to the inter-photon entanglement between them. Thus, the protocol differs from the usual swapping protocols in the literature and also from \cite{adhikari2010}. 

The same setup can also be used to create path-path and path-polarization hybrid entanglement between the two photons. For this, a few small changes need to be adequately incorporated in the setup. The sketch of this is as follows: If one wants the path-path or path-polarization intra-photon entanglement by performing small changes from the current setup one can start from Eq.(\ref{joint}). For example, by blocking $d_2$ and $d_7$ one gets a state (un-normalized) $(\alpha \gamma \hat{d}^{\dagger}_{1H}\hat{d}^{\dagger}_{5H}-\beta \delta \hat{d}^{\dagger}_{1V}\hat{d}^{\dagger}_{5V} )$. Now, introduce a $PBS$ along the path $d_1$ which transmits horizontal and reflects vertical polarization. Let the state of the photon along the output mode of $PBS$ are $d'_{1H}$ and $d''_{1V}$ and a polarization flipper is used along the path $d''_{1}$. This provides a path-polarization hybrid entanglement $|\psi'_{12}\rangle = N_3 (\alpha \gamma \hat{d'}^{\dagger}_{1H}\hat{d}^{\dagger}_{5H}-\beta \delta \hat{d''}^{\dagger}_{1H}\hat{d}^{\dagger}_{5V} )$, where $N_3$ is the normalization constant. Similar to operations taken along path $d_1$, if one suitably does those operations along path $d_5$ then a path-path intra-photon entanglement can be produced is given by  $|\psi''_{12}\rangle = N_3 (\alpha \gamma \hat{d'}^{\dagger}_{1H}\hat{d'}^{\dagger}_{5H}-\beta \delta \hat{d''}^{\dagger}_{1H}\hat{d''}^{\dagger}_{5H} )$ where $d'_{5}$ and $d''_{5}$ are output mode of $PBS$ along $d_5$. 

\vspace{2cm}
\section{Quantum state transfer}
As mentioned before, our setup can also be used for demonstrating the teleportation of an unknown quantum state. One may say that it is an obvious fact that once we have generated the entangled state $|\Psi_{AB}\rangle$, the teleportation is one more step. For this, one more qubit needs to be brought either by Alice or Bob followed by a relevant Bell-basis measurement.  However, it seems interesting if the polarization state belonging to Alice to Bob can be teleportated without introducing another qubit state and Bell-basis analysis. We exactly provide such a scheme of state transfer.

In order to demonstrate such a state transfer protocol, let us use two polarization rotators $PR_4$ and $PR_5$ along the spatial modes $d_5 $ and $d_7 $ respectively. So that, the creation-operators $\hat{d}^{\dagger}_{5H}$ and  $\hat{d}^{\dagger}_{5V}$ transformed as,  $\hat{d}^{\dagger}_{5H}=\frac{1}{\sqrt{2}}(\hat{d}^{\dagger}_{5H}+\hat{d}^{\dagger}_{5V})$ and  $\hat{d}^{\dagger}_{5V}=\frac{1}{\sqrt{2}}(\hat{d}^{\dagger}_{5H}-\hat{d}^{\dagger}_{5V})$ and similarly for $\hat{d}^{\dagger}_{7H}$ and $\hat{d}^{\dagger}_{7V}$. After these two rotations, the state given by Eq.(\ref{joint}) can be written as
\begin{widetext}
\begin{eqnarray}
\nonumber
	|\Psi_{5} \rangle &=& \frac{N_2}{2 \sqrt{2}}[(\alpha \gamma \hat{d}^{\dagger}_{1H}-\beta \delta \hat{d}^{\dagger}_{1V})\hat{d}^{\dagger}_{5H}+(\alpha \gamma \hat{d}^{\dagger}_{1H}+\beta \delta \hat{d}^{\dagger}_{1V})\hat{d}^{\dagger}_{5V}\\ 
			\nonumber
			&+&i((\alpha \gamma \hat{d}^{\dagger}_{2H}+\beta \delta \hat{d}^{\dagger}_{2V})\hat{d}^{\dagger}_{5H}+(\alpha \gamma \hat{d}^{\dagger}_{2H}-\beta \delta \hat{d}^{\dagger}_{2V})\hat{d}^{\dagger}_{5V})\\ 
			\nonumber
			&+&i((\alpha \gamma \hat{d}^{\dagger}_{1H}+\beta \delta \hat{d}^{\dagger}_{1V})\hat{d}^{\dagger}_{7H}+(\alpha \gamma \hat{d}^{\dagger}_{1H}-\beta \delta \hat{d}^{\dagger}_{1V})\hat{d}^{\dagger}_{7V} )\\ 
			\nonumber
			&-&((\alpha \gamma \hat{d}^{\dagger}_{2H}-\beta \delta \hat{d}^{\dagger}_{2V})\hat{d}^{\dagger}_{7H}+(\alpha \gamma \hat{d}^{\dagger}_{2H}+\beta \delta \hat{d}^{\dagger}_{2V})\hat{d}^{\dagger}_{7V})]|0 \rangle
			 \\
\end{eqnarray}
 \end{widetext}
After $PR_{4}$ and $PR_{5}$ operations, Bob uses two polarizing beam splitters, $PBS_{4}$ and $PBS_{5}$ along the modes $d_5$ and $d_7$ and detects the photons in four detectors $D_5$, $D_{6}$, $D_{7}$ and $D_{8}$. For four outcomes of Bob yield eight different possibilities at Alice's end. The states of the Bob's photon corresponding to the detectors $D_5$, $D_{6}$, $D_{7}$ and $D_{8}$ are $\hat{d}^{\dagger}_{5H}|0 \rangle$, $\hat{d}^{\dagger}_{5V}|0 \rangle$, $\hat{d}^{\dagger}_{7H}|0 \rangle$ and $\hat{d}^{\dagger}_{7V}|0 \rangle$ respectively. The measurements at Bob's end thus produce the following states unnormalized at Alice's end are given by
\begin{subequations}
\begin{eqnarray}
|\Psi_{D5} \rangle &=& [(\alpha \gamma \hat{d}^{\dagger}_{1H}-\beta \delta \hat{d}^{\dagger}_{1V})\\
\nonumber
&+& (\alpha \gamma \hat{d}^{\dagger}_{2H}+\beta \delta \hat{d}^{\dagger}_{2V})]|0 \rangle \\
 |\Psi_{D6} \rangle&=&[(\alpha \gamma \hat{d}^{\dagger}_{1H}+\beta \delta \hat{d}^{\dagger}_{1V})\\
 \nonumber
 &+&  ( \alpha \gamma \hat{d}^{\dagger}_{2H}-\beta \delta \hat{d}^{\dagger}_{2V}) ]|0 \rangle\\
|\Psi_{D7} \rangle&=&  [(\alpha \gamma \hat{d}^{\dagger}_{1H}+\beta \delta \hat{d}^{\dagger}_{1V})   \\
\nonumber
&+&(\alpha \gamma \hat{d}^{\dagger}_{2H}-\beta \delta \hat{d}^{\dagger}_{2V})]|0 \rangle \\
 |\Psi_{D8} \rangle&=&[(\alpha \gamma \hat{d}^{\dagger}_{1H}-\beta \delta \hat{d}^{\dagger}_{1V})  \\
 \nonumber
 &+& (\alpha \gamma \hat{d}^{\dagger}_{2H}+\beta \delta \hat{d}^{\dagger}_{2V}) ]|0 \rangle
 \end{eqnarray}
\end{subequations}
Note here that $|\Psi_{D_5} \rangle=|\Psi_{D_8} \rangle$ and $|\Psi_{D_6} \rangle=|\Psi_{D_7} \rangle$. Let us now assume that $\alpha = \beta = 1/\sqrt{2}$. After the detection of photon in four different detectors ($D_5,D_6,D_7$ and $D_8$), Bob needs to send the information through a classical communication channel. Following Bob's instruction, Alice performs suitable gate operations to obtain the desired polarization state $|\psi'_3 \rangle = (\gamma \hat{d}^{\dagger}_{1H}+ \delta \hat{d}^{\dagger}_{1V})|0 \rangle$ of Bob as given in the Table-I.
\begin {table}
\begin{center}
\begin{tabular}{|p{2.5 cm} | c| c|}
\hline 
\multirow{2}{*}{Bob's detection} & \multicolumn{2}{|c|}{Alice's operation} \\
 \cline{2-3}
   & on $d_{1}$ & on  $d_{2}$\\
\hline
\centering
 $ D_5$   &  $\hat{\sigma}_{Z}$    & $\hat{\mathbb{I}}$ \\
\hline
\centering
$ D_6 $ & $\hat{\mathbb{I}}$  &  $ \hat{\sigma}_{Z}$ \\
\hline
 \centering
 $ D_7 $   & $\hat{\mathbb{I}}$ &  $ \hat{\sigma}_{Z} $\\
\hline
\centering
 $ D_8 $ &  $ \hat{\sigma}_{Z} $   & $\hat{\mathbb{I}}$ \\
\hline
\end{tabular}
\caption{Alice's unitary rotation along the path modes $d_{1}$ and  $d_{2}$ upon receiving instructions from Bob. }
\end{center}
\end{table}
Whenever Bob detects photon in $D_5$ or in $ D_{8}$, he asks Alice to use a Pauli gate $\hat{\sigma_{z}}$ in the mode $d_{1} $. If he gets the photon in $ D_{6} $ or in  $ D_{7} $, Alice has to use the $\hat{\sigma_{z}} $ in the mode $d_{2}$. Hence, we demonstrated a state transfer protocol from Bob to Alice without any direct interaction between photons in two interferometers $MZ_{1}$ and $MZ_{3}$. Note that the success probability of teleportation in this case is $1/8$, i.e., the cost of the state transfer is larger than the original teleportation protocol. Importantly, no Bell-basis measurement is required in the whole process.
\section{Discussion}
We have demonstrated an interesting swapping protocol using simple linear optical devices where the intra-photon entanglement between path and polarization degrees of freedom of a single photon is swapped to polarization-polarization entanglement of two spatially separated photons. Note that, those photons have never interacted during the whole process. We have further shown how the same setup can be used for the purpose of a curious quantum state transfer. Both the protocols avoid Bell basis discrimination which is taken care by exploiting the actions of the spatial modes in $MZ_1$ and $MZ_3$. We believe that the proposed setup can be experimentally implemented with the existing technology that uses  linear optical devices.

\section*{Acknowledgments}
AG and MLB acknowledge the local hospitality of NIT Patna during their visits and financial support from Ramanujan Fellowship research grant (SB/S2/RJN-083/2014). AKP acknowledges the project DST/ICPS/QuEST/2018/Q-42.

\end{document}